\documentstyle[multicol,aps,prl,floats]{revtex}
\draft

\begin{document}
\draft
\title{Polymer Stretching by Turbulence}
\author{Michael Chertkov}
\address{Theoretical Division, T-13 \& CNLS\\
Los Alamos National Laboratory,\\
Los Alamos, NM 87545}
\date{Submitted to PRL November 8, 1999; resubmitted February 22, 2000}
\maketitle

\begin{abstract}
The stretching of a polymer chain by a large scale chaotic flow is
considered. The steady state which emerges as a balance of the turbulent
stretching and anharmonic resistance of the chain is quantitatively
described, i.e. the dependency on the flow parameters (Lyapunov exponent
statistics) and the chain characteristics (the number of beads and the
inter-bead elastic potential) is made explicit. 
%Implications for the drag reduction theory are discussed.
\end{abstract}

\pacs{PACS numbers 47.10.+g, 47.27.-i, 05.40.+j, 61.25.H}

\begin{multicols}{2} 

A great task is to develop a theory of turbulence for a dilute polymer
solution which explains, particularly, polymer drag reduction, the so-called Toms' effect 
\cite{49Tom}. Although a very plausible qualitative theory of polymer's drag
reduction, describing the bulk viscosity and the size of the viscous boundary
layer increase was given by Lumley \cite{69Lum,72Lum}, no quantitative
theory of the phenomenon is available.
It is also widely accepted that the enormous increase of the polymer
contribution into the stress tensor (by factor of ten thousand in a
turbulent flow, e.g. \cite{77Hin}, and
up to $130$ times in the recent
experiment of Groisman and Steinberg on elastic turbulence at very low
Reynolds number \cite{99GS})) is associated with
extreme polymer stretching \cite{69Lum,77Hin} %,87BCAH,95GB}
(there exists an alternative theory suggesting that the connection between
the turbulent strain rates and the elongational viscosity is circumstantial 
\cite{86TD}, see also a comprehensive discussion of the subject in \cite
{99SW}). Therefore, it seems logical to start a theoretical exploration
of the whole problem of the polymer solution hydrodynamics from
description of a single polymer stretching.

In the present letter I discuss a polymer chain placed in a
statistically known random flow under the assumption that the back reaction
of the polymer on the flow is negligible, i.e. the polymer is
passive. 
Polymer stretching in laminar flows was extensively studied in the past, e.g. 
\cite{77Hin}. A recent
experimental advance has come with the state-of-the-art capability to
visualize a polymer (DNA molecule) and therefore to study its stretching
directly \cite{95Chu,99Chu}. 
A qualitative theory of polymer stretching by chaotic flow
\cite{72Lum} has predicted the jump of the mean square polymer length
at $O(1)$ value of the stretching-to-linear elasticity ratio.
Based on the
recent progress in the theory of the passive scalar advection by a
large-scale flow \cite{59Bat,68Kra,94SS,95CFKLa,94CGK,BCKL95,95SS,97BGK,98CFK,99BF} 
Balkovsky et al. \cite{99BFL} have developed the approach of
\cite{72Lum} constructed
a quantitative theory of the transition onset. 
In the present letter I generalize the approach of \cite{99BFL}.
Nonlinear, 
but still passive theory of polymer stretching above
as well as below the aforementioned threshold is delivered .
Of special note is  the described 
dependence on the equilibrium polymer contour length.

Equilibrium and even extended-by-flow lengths of a polymer chain are
typically much smaller than the Kolmogorov viscous scale. This allows to
model a velocity field (counted from a reference point, say one of the
polymer's end points) advecting the polymer by $u^{\alpha }(t;\rho)=\sigma
^{\alpha \beta }(t)\rho^{\beta }$ (i.e. the velocity at all scales under
consideration changes linearly with the separation). Here the statistics of
the traceless $\hat{\sigma}$ (we consider incompressible flow) is assumed to
be known. I will analyze a sequence of models moving steadily from  (\ref
{edbseq}), introduced in \cite{99BFL}, through (\ref{undb}) to (\ref
{unchbound}). Polymers will be modeled by a chain of beads (two beads in the
cases of (\ref{edbseq}) and (\ref{undb}) and $N$ beads for (\ref{unchbound}%
)) connected by elastic (harmonic for (\ref{edbseq}) and anharmonic for (\ref
{undb}) and(\ref{unchbound})) springs. Forces of friction acting on a bead,
which occurs as the bead is moving against the local flow, are equilibrated
by elastic tension within the springs attached to the bead and by thermal
forces. The goal is to describe statistics of the passive polymer(s)
advected by ${\bf u}(t;r)$.

I start by elaborating on the features and defects of the
two-bead-one-linear-spring model. The basic equation of motion for the
inter-bead separation ${\bf \rho }$ is 
\begin{equation}
\frac{d}{dt}\rho ^{\alpha }-\sigma ^{\alpha \beta }(t)\rho ^{\beta }=-k\rho
^{\alpha }+\xi ^{\alpha }(t),  \label{edbseq}
\end{equation}
where the friction force on the rhs of (\ref{edbseq}) is equated by elastic
and thermal forces on the lhs. In (\ref{edbseq}) $k$ is the ratio of the
string linear tension to the friction coefficient; $\left\langle \xi (0)\xi
(t)\right\rangle =2\kappa \delta \left( t\right) $ , and $\kappa $ is the
thermal (molecular) diffusion coefficient. The solution of (\ref{edbseq}) is
given by 
\begin{equation}
{\bf \rho }(t)=e^{-kt}\hat{W}(t)\left[ {\bf \rho }(0)+\int%
\limits_{0}^{t}e^{kt^{\prime }}\hat{W}^{-1}(t^{\prime }){\bf \xi }(t^{\prime
})\right] ,  \label{W}
\end{equation}
with $\hat{W}(t)$ is the time-ordered exponential $T\exp \left[
\int\limits_{0}^{t}dt^{\prime }\hat{\sigma}(t^{\prime })\right] $. The
temporal evolution of $\hat{W}$ is described by two Lyapunov exponents at
times greater than the inverse of the exponents, while all other
(''angular'') degrees of freedom are frozen \cite{98CFK}. Respectively, $tr[%
\hat{W}\hat{W}^{+}]$ grows as $\exp [2$ $\lambda t]$ and one gets the
following estimation for the length of the polymer, $\rho ^{2}(t)\sim \max
\left\{ 1,\exp [2(\lambda -k)t]\right\} \kappa /\left| \lambda -k\right| $.
Here the long-time statistics of the leading Lyapunov exponent (according to
a matrix version of the large deviation theory \cite{99BF}) is described by
the weight 
\begin{equation}
d\lambda \exp \left[ -tS(\lambda )\right] .  \label{lambdaWeight}
\end{equation}
The entropy function $S(\lambda )$ is concave with a zero 
minimum at $\lambda =%
\bar{\lambda}$. 
%( normalization requires $S(\bar{\lambda})=0$ ) . 
At the largest times, the moments of $\rho $ 
are governed by a saddle point value, $%
\lambda =\lambda _{n}$ , $S^{\prime }\left( \lambda _{n}\right) =n$, 
\begin{equation}
\frac{\left\langle \rho (t)^{n}\right\rangle }{\kappa ^{n/2}}\sim \max
\left\{ \frac{1}{\left| \bar{\lambda}-k\right| ^{n/2}},\frac{e^{t\left[
-nk+n\lambda _{n}-S\left( \lambda _{n}\right) \right] }}{\left| \lambda
_{n}-k\right| ^{n/2}}\right\} .  \label{Slan}
\end{equation}
This result shows, particularly, that only the lowest moments (and only if $%
k>\bar{\lambda}$ ) are steady. However small the $\bar{\lambda}$ (in
comparison with $k$ ) may be it can not prevent the higher moments from an
asymptotic in time growth, the steady PDF of $\rho $ decays algebraically 
\cite{99BFL} with a large-$\rho $-cutoff growing in time. The threshold for
the $n$-th moment convergence is given by 
\begin{equation}
S^{\prime }\left( \lambda _{n}\right) =n,\qquad k_{n}=\lambda _{n}-S\left(
\lambda _{n}\right) /n.  \label{threshold}
\end{equation}
$k_{n}$ as a function of $n$ has a minimum, $\bar{\lambda}$ , at $n=0$ . 
The  $n=2$ value of the threshold is associated with the convergence of the
dilute polymer solution contribution into the overall strain rate, and
therefore can be identified as  the onset of the drag reduction \cite{99BFL}. 
The dependence of the onset value on the number of elements in a long
polymer chain (i.e. on square equilibrium radius of gyration of the polymer) will be
discussed later in the letter. 
A problem, however, is that 
the linear theory does not allow to describe
statistics of polymer (and, particularly, 
to find the steady length of the stretched
polymer) above the drag reduction onset.

Account for anharmonic elasticity, 
we are switching our attention to, solves the latter
problem. The anharmonic generalization of (%
\ref{edbseq}) to consider is 
\begin{equation}
\frac{d}{dt}\rho ^{\alpha }-\sigma ^{\alpha \beta }(t)\rho ^{\beta
}=-\partial _{\rho }^{\alpha }U\left( \rho \right) +\xi ^{\alpha }(t),
\label{undb}
\end{equation}
where $U(\rho )$ is an arbitrary bounded potential. 
Isotropy dictates that the distribution function
of $\rho ^{\alpha }$ should depend on the polymer length only. For the 
dynamically relevant times, $t\gg 1/\bar{\lambda}$, the
problem reduces to a study of (\ref{undb}) projected on the major stretching
direction. First, consider the case of the Gaussian entropy function, $%
S(\lambda )=(\lambda -\bar{\lambda})^{2}/2\Delta $, for which the reduction
from the stochastic differential equation (\ref{undb}) to the Fokker-Planck
equation for the distribution function of the separation $\rho $, ${\cal P}$, 
is especially simple 
\begin{equation}
\left\{ \partial _{t}+\frac{\Delta }{2}\partial _{\rho }\rho \partial _{\rho
}\rho -\bar{\lambda}\partial _{\rho }\rho +2\kappa \partial _{\rho
}^{2}+\partial _{\rho }U^{\prime }(\rho )\right\} {\cal P}=0.  \label{P}
\end{equation}
The stationary solution of (\ref{P}) is 
\begin{equation}
{\cal P}(\rho )=C\rho ^{-1}\exp \left[ -\int\limits_{0}^{\rho }dr^{\prime }%
\frac{U^{\prime }(r^{\prime })-\bar{\lambda}r^{\prime }}{2\kappa +\Delta
r^{\prime 2}/2}\right] ,  \label{Pstat}
\end{equation}
where $C$ is the normalization coefficient. One observes that when the
molecular diffusivity is small it enters only as an ultraviolet cutoff, $%
r_{d}=\sqrt{\kappa /\Delta }$ in the large scale ($\rho \gg r_{d}$ )
behavior. In the forthcoming calculations the diffusion will be
considered small. We will not be counting for effect of diffusion explicitly
and yet introducing the $r_{d}$ cutoff if appropriate. The general case of a
non-Gaussian entropy function does not lead to a simple differential
equation for the PDF. However, it may always be described by the path
integral 
\begin{eqnarray}
{\cal P}\left( r\right)  &=&\int\limits^{\rho (0)=r}{\cal D}\lambda (t){\cal %
D}\rho (t){\cal D}p(t)\exp \left[ -\int\limits_{-\infty }^{0}{\cal L}dt%
\right] ,  \label{Pr0} \\
{\cal L} &\equiv &-p\left[ \dot{\rho}-\lambda \rho +U^{\prime }\left( \rho
\right) \right] +S\left( \lambda \right) .  \label{Lpr}
\end{eqnarray}
The effective action in (\ref{Pr0}) is large as the time of evolution in
the strong flow is. Therefore, the semi-classical saddle-point approximation
should work. The saddle system of equations, supplied by the final condition
(at the observation time) $\rho (0)=r$ , is 
\begin{equation}
\left\{ 
\begin{array}{c}
\dot{\rho}_{sp}-\lambda _{sp}\rho _{sp}+U^{\prime }\left( \rho _{sp}\right)
=0, \\ 
-\dot{p}_{sp}-\lambda _{sp}p_{sp}+p_{sp}U^{\prime \prime }\left( \rho
_{sp}\right) =0, \\ 
p_{sp}\rho _{sp}+S^{\prime }\left( \lambda _{sp}\right) =0.
\end{array}
\right.   \label{spsys}
\end{equation}
Notice also that this system of equations is consistent with the
conservation of energy 
\begin{equation}
E=-p_{sp}\left[ -\lambda _{sp}\rho _{sp}+U^{\prime }\left( \rho _{sp}\right) 
\right] +S\left( \lambda _{sp}\right) =const.  \label{Econst}
\end{equation}
Because one is looking for finite action during an infinite time (beginning
at $t=-\infty $ ) evolution, the choice $E=0$ is mandatory. The saddle point
PDF is given by 
\begin{equation}
{\cal P}_{sp}(r)\sim \exp \left[ -\int\limits_{0}^{r}\frac{d\rho _{sp}}{\rho
_{sp}}S^{\prime }\left( \lambda _{sp}\right) \right] ,  \label{Psp1}
\end{equation}
with $\lambda _{sp}$ and $\rho _{sp}$ related to each other locally through $%
U^{\prime }(\rho _{sp})/\rho _{sp}=\lambda _{sp}-S\left( \lambda
_{sp}\right) /S^{\prime }\left( \lambda _{sp}\right) $ . For the Gaussian $%
S(\lambda )$, one derives $\dot{\rho}_{sp;G}+\bar{\lambda}\rho
_{sp;G}-U^{\prime }\left( \rho _{sp;G}\right) =0$ and 
\begin{equation}
{\cal P}_{sp;G}(r)\sim \exp \left[ {\cal -}\frac{2}{\Delta }%
\int\limits_{r_{d}}^{r}\frac{dr^{\prime }}{r^{\prime 2}}\left( U^{\prime
}(r^{\prime })-\bar{\lambda}r^{\prime }\right) \right] .  \label{LspG}
\end{equation}
Comparing the semi-classical expression (\ref{LspG}) with the exact one (\ref
{Pstat}), one concludes that the algebraic $1/r$ factor is gained through
the fluctuations calculated on the top of the saddle solution. (Notice also,
that the saddle-point solution becomes exact after changing to log $\ln
[r/r_{d}]$ variables). To find the steady length of the polymer, one should
also perform an additional variation of the effective action in (\ref{LspG})
with respect to $r$. The result is given by the time-independent solution of
(\ref{spsys}) 
\begin{equation}
\bar{\lambda}\rho _{st}=U^{\prime }\left( \rho _{st}\right) .  \label{rhost}
\end{equation}
The last equation is valid for an arbitrary form of $S(\lambda )$.

\bigskip Consider now the case of a long polymer modeled
by  $N$ anharmonic springs connected in a chain 
\begin{equation}
\frac{d}{dt}\rho _{i}^{\alpha }\!=\!\sigma ^{\alpha \beta }\rho _{i}^{\beta
}\!-\!\left[ \partial ^{\alpha }U\left( {\bf \rho }_{i;i-1}\right)\!+\!%
\partial ^{\alpha }U\left( {\bf \rho }_{i;i+1}\right) \right]\!+\!\xi
_{i}^{\alpha },  \label{unchbound}
\end{equation}
with ${\bf \rho }_{i;k}\equiv {\bf \rho }_{i}-{\bf \rho }_{k}$ , and the
boundary condition, ${\bf \rho }_{1;0}={\bf \rho }_{N+1;N}=0$ , imposed at
the fictitious end points of the polymer, added for convenience. As the
strong flow regime when the chain is extremely stretched (so that
anharmonicity becomes more important while off-linear conformations of the
polymer are less relevant) one can 1) take into account only linear
(straight-line) conformations of the polymer; 2) neglect diffusion; 3)
replace the projected (on the major stretching direction) $\hat{\sigma}$ by
the leading Lyapunov exponent. The continuous version of the scalar
projection of (\ref{unchbound}) (to be considered for $N\gg 1$, the major
case of interest) is 
\begin{equation}
\frac{d}{dt}\rho _{n}=\lambda (t)\rho _{n}+\partial _{n}U^{\prime }\left(
\partial _{n}{\bf \rho }\right) ,\qquad  \label{rhon}
\end{equation}
supplemented by $\partial _{n}{\bf \rho }\left( 0\right) {\bf =}\partial _{n}%
{\bf \rho }\left( N\right) =0$ . 
%Averaging with respect to the $\lambda (t)$
%statistics is assumed, obviously leading to the generalization of
%path integral (\ref{Pro},\ref{Lpr}) for the case of many particles (beads).
The saddle-point method explained above in detail for the single-spring
model is also valid (it is even enhanced as $N$ appears to be an additional
saddle parameter) for analyzing the following system of equations 
\begin{equation}
\left\{ 
\begin{array}{c}
\dot{\rho}_{sp}-\lambda _{sp}\rho _{sp}+\partial _{n}U^{\prime }\left(
\partial _{n}{\bf \rho }_{sp}\right) =0, \\ 
-\dot{p}_{sp}-\lambda _{sp}p_{sp}+\partial _{n}^{2}p_{sp}U^{\prime \prime
}\left( \partial _{n}{\bf \rho }_{sp}\right) =0, \\ 
\int\limits_{0}^{N}p_{sp}\rho _{sp}dn+S^{\prime }\left( \lambda _{sp}\right)
=0.
\end{array}
\right.  \label{spsystn}
\end{equation}
The zero energy condition, consistent with (\ref{spsystn}), is $%
\int\limits_{0}^{N}dnp_{sp}\left[ -\lambda _{sp}\rho _{sp}+\partial
_{n}U^{\prime }\left( \partial _{n}\rho _{sp}\right) \right] =S\left(
\lambda _{sp}\right) $ . The saddle point system of equations supplied by
the final condition $\rho _{n}(t=0)=r(n)$ has a unique solution. The
functional PDF is given by, ${\cal P}\left\{ r\left( n\right) \right\} \sim
\exp \left[ \int_{-\infty }^{0}p_{sp}\dot{\rho}_{sp}dt\right] $. 
If, however, one
is interested in the PDF of the final point (polymer length) only, variation
with respect to all the $r(n)$ except the end of the polymer, $r(N)$ ,
should be taken. A very symmetric solution of the enhanced saddle point
system is available. The hint is that not only the global energy vanishes
but also the energy density (for each and every value of $n$) does. This
leads to 
\begin{equation}
\lambda _{sp}-\frac{\partial _{n}U^{\prime }(\partial _{n}\rho _{sp})}{\rho
_{sp}}=\frac{S(\lambda _{sp})}{S^{\prime }(\lambda _{sp})}=\frac{\dot{\rho}%
_{sp}}{\rho _{sp}}.  \label{uniform solution}
\end{equation}
The first equation in (\ref{uniform solution}) can be integrated 
\[
n+\left( N-2n\right) \theta \left( \frac{N}{2}-n\right) =\frac{%
\int\limits_{0}^{\tau}\frac{U^{\prime \prime }(x)dx}{\sqrt{2\int_{x}^{\tau
_{\ast }}yU^{\prime \prime }(y)dy}}}{\sqrt{\lambda _{sp}-\frac{S(\lambda
_{sp})}{S^{\prime }(\lambda _{sp})}}},
\]
where $\tau\equiv \partial_n \rho_{sp}$, and $\theta (x)$ is the step
function. Therefore, the maximal tension on the string $\tau _{\ast }(t)$
and the length of the polymer, $\rho _{\ast }(t)=\rho _{sp}(t;N)$, are
related to each other through 
\begin{eqnarray}
N\sqrt{\lambda _{sp}-\frac{S(\lambda _{sp})}{S^{\prime }(\lambda _{sp})}} &=&%
\sqrt{2}\int\limits_{0}^{\tau _{\ast }}\frac{U^{\prime \prime }(x)dx}{\sqrt{%
\int_{x}^{\tau _{\ast }}yU^{\prime \prime }(y)dy}},  \label{latau} \\
\frac{\rho _{\ast }^{2}}{4}\left[ \lambda _{sp}-\frac{S(\lambda _{sp})}{%
S^{\prime }(\lambda _{sp})}\right] &=&\int\limits_{0}^{\tau _{\ast
}}xU^{\prime \prime }(x)dx.  \label{rhotau}
\end{eqnarray}
Finally, according to the second part of (\ref{uniform solution}), applied
to $n=N$ , the dynamics of all the time-dependent fields ( $\lambda_{sp}(t)$%
, $\rho _{\ast }(t)$ , $\tau _{\ast }(t)$ ) is unambiguously fixed by $d\ln 
\left[ \rho _{\ast }(t)\right] /dt=S(\lambda _{sp})/S^{\prime }(\lambda
_{sp})$ , $\rho _{\ast }(0)=R$ . The probability to observing a polymer of
length $R$ is given by (\ref{Psp1}) with $\rho _{sp}$ replaced by $\rho
_{\ast }$, $\rho _{\ast }$ and $\lambda _{sp}$ related to each other through
(\ref{latau},\ref{rhotau}). The PDF has a pronounced maximum at some value, $%
R_{\ast }$ , which is the steady state polymer length. $R_{\ast }$ is
described by (\ref{latau},\ref{rhotau}) taken at $\lambda _{sp}\rightarrow 
\bar{\lambda}$ (then $\rho _{\ast }\rightarrow R_{\ast }$ ).

The algebraic system of equations (\ref{latau},\ref{rhotau}), with the
result substituted in the $\rho _{\ast }\rightarrow \rho _{sp}$ version of (%
\ref{Psp1}), solves unambiguously the problem of finding the probability for
the $N$-chain polymer characterized by the inter-bead interaction $U(x)$ to
be stretched to the length $R$ in the flow described by $S(\lambda )$. For
the Gaussian entropy function $S\left( \lambda \right) $ and the algebraic
potential $U(x)=qx^{2+\alpha }$ one derives 
\begin{eqnarray}
&&{\cal P}\left( R\right) \sim \left( \frac{R}{r_{d}}\right) ^{2\bar{\lambda}%
/\Delta }\exp \left[ -\frac{2\bar{\lambda}}{(2+\alpha )\Delta }\left( \frac{R%
}{R_{\ast }}\right) ^{2+\alpha }\right] ,  \label{PRch} \\
&&R_{\ast }=\frac{2\left[ \bar{\lambda}/q\right] ^{1/\alpha }}{\left(
1+\alpha \right) ^{1/\alpha }}\left( \frac{\Gamma \left[ \frac{4+3\alpha }{%
4+2\alpha }\right] }{\Gamma \left[ \frac{1+\alpha }{2+\alpha }\right] \sqrt{%
2\pi }}N\right) ^{1+2/\alpha }.  \label{R*}
\end{eqnarray}
Notice also, that (\ref{R*}) holds for an arbitrary form of $S(\lambda )$.
For the linear harmonic chain, $U=kx^{2}$ , and the Gaussian entropy
function, one gets, $\Delta \ln \left[ P(R)\right] /\left( 2\ln
[R/r_{d}]\right) =\bar{\lambda}-\pi ^{2}k/N^{2}$ , i.e. for $N$ greater then 
$\pi \sqrt{\bar{\lambda}/k}$ the pure harmonic consideration has no sense as
the resulted pdf is not normalizable. In other words, a transition to
the nonlinear regime is taking place at a very small ($\pi ^{2}k/N^{2}$) value
of $\bar{\lambda}$.

Knowledge of $P(R)$ allows to calculate any correlation functions of $R$. I
will briefly discuss two examples of interest. First of all, as it was
already mentioned above, the kinetic theory estimate for the polymer
contribution into strain tensor is proportional to $\left\langle
R^{2}\right\rangle \equiv \int R^{2}P(R)dR$ \cite{99BFL}. Therefore, (%
\ref{latau},\ref{rhotau},\ref{Psp1}) applied to the case of a complex
potential $U(x)=kx^{2}+qx^{2+\alpha }$, will describe the ''first order
phase transition'' (jump of the $\left\langle R^{2}\right\rangle $ value)
from the linear theory estimate, $\sim r_{d}^{2}$, valid at $\bar{\lambda}%
+2\Delta <\pi ^{2}k/N^{2}$, upto $\sim R_{\ast }^{2}$, correspondent to the
nonlinear case.  Second example explains how to
recalculate the probability, $P_{\deg }$, of the polymer to degrade (to
break in two pieces) under the action of turbulence. Indeed, assume that the
polymer string breaks if the maximal (along the polymer) tension exceeds the
critical value, $\tau _{c}$. Then, $P_{\deg }$ is given by the integration
of $P(R)$ from $R_{c}$ upto $\infty$, where $R_{c}$ solves the system (\ref
{latau},\ref{rhotau}) with $\rho _{\ast }$ and $\tau _{\ast }$ replaced by $%
R_{c}$ and $\tau _{c}$ respectively.

\bigskip 
%The models considered may be imperfect from the point of view of
%potential applications. 
Let me briefly mention other passive problems of interest 
which remained beyond the scope
of the present letter. First of all, it is known that
in the case of fully developed turbulence,
the degenerate (or very close to degenerate) sheet-like flow configurations
(associated with the so-called Vieillefosse tail of the velocity gradient
PDF) can be favored, especially in the viscous range \cite{99CPS}. 
Polymer stretching in the
degenerate case requires a separate investigation accounting for
off-straight-line but planar conformations of the polymer. Notice also, that
turbulent stretching of a flexible but inextensible polymer \cite{77Hin}
(the one which is considered to be a good model of B-DNA \cite{95MS}) is yet
another problem which requires an accurate account for the off-straight-line
conformations. %Second, the case of a
%moderate (and/or small) value of advection-to-diffusion ratio is also an
%interesting one to study. One should be able to describe quantitatively the
%transition from the Gaussian coil, described within the equilibrium
%statistical theory of polymers \cite{86DE,94GK}, to the extremely stretched
%state, studied here, with a gradual increase of the flow intensity. 
Second, excluded volume effect along with hydrodynamic self-interaction of
the polymer, known to be of a great importance in the equilibrium theory 
\cite{86DE,94GK}, may also play a significant role in the random flow case.
Third, compressibility of the flow (responsible for the sign
change of $\bar{\lambda}$ \cite{98CKV}) may cause new effects interesting
enough to study (the saddle point approach will not work if $\bar{\lambda}$
is negative). Finally, steady statistics of comb-, star-like and randomly
branched polymers, dendrimers or polymeric membranes, placed in a chaotic
flow can also be studied within the passive approach demonstrated above for
the linear polymer.

If concentration of polymer
solution is small enough the passive approach developed above is valid.
However, the elastic contribution into the strain tensor grows 
with the concentration. Once the elastic contribution into the stress tensor 
becomes of the order of the kinetic one
the problem cannot be treated as passive. 
The steady length of the polymer 
%at the edge of the transition to the active regime 
may be still much smaller than the viscous scale. 
The scale separation, in principle, allows to generalize the passive 
approach of the present work and to describe various chaotic effects 
of the non-Newtonian hydrodynamics 
(see \cite{99CheB} for an attempt to proceed in the active direction).

Very fruitful discussions with E. Balkovsky, L. Burakovsky,
A. Groisman, S. Tretiak and D. Preston are gratefully acknowledged. 
I am also thankful to the authors of 
\cite{99SW,99BFL} for sending their preprints prior to publication and to G.
Falkovich and G. Doolen for valuable comments. This work was supported by a
J. R. Oppenheimer fellowship.

\end{multicols}

\end{document}